\documentclass[12pt]{article}
\usepackage{amsfonts}
\usepackage{amsmath}
\usepackage{amssymb}
\usepackage{amscd}
\usepackage[dvips]{graphicx}

\begin{document}

\begin{center}
{\Large \textbf{Braided Field Quantization from Quantum Poincare Covariance}}
\end{center}

\bigskip

\begin{center}
{{\large ${\mathrm{Jerzy\;Lukierski}}$, ${\mathrm{Mariusz\;Woronowicz}}$ }}

\bigskip

{
$\mathrm{~Institute\;of\;Theoretical\;Physics}$}

{
$\mathrm{\ University\; of\; Wroclaw\; pl.\; Maxa\; Borna\; 9,\; 50-206\;
Wroclaw,\; Poland}$}

{
$\mathrm{\ e-mail:\;lukier@ift.uni.wroc.pl;woronow@ift.uni.wroc.pl}$}

\bigskip
\end{center}

\begin{abstract}
We demonstrate that the covariance of the algebra of quantum NC fields under
quantum-deformed Poincare symmetries implies the appearence of braided
algebra of fields and the notion of braided locality in NC QFT. We briefly
recall the historical development of NC QFT which was firstly formulated in
the framework using classical relativistic symmetries but further it was
described as generated by the quantum-deformed symmetries. We argue that
consistent covariant quantum-deformed formalism requires "braiding all the
way", in particular braided commutator of deformed field oscillators as well
as the braid between the field oscillators and noncommutative Fourier
exponentials. As example of braided quantum-deformed NC QFT we describe the
NC scalar free fields on noncommutative canonical (Moyal-Weyl) space-time
with braided $c$-number field commutator which implies braided locality.
\end{abstract}

\section{Introduction}

It is believed that due to quantum gravity (see e.g. \cite{lit1}) or
quantized string effects (see \cite{lit4}) one should investigate QFT on
noncommutative space-times. In such new field-theoretic models the standard
Minkowski space-times coordinates $x_{\mu }$ are replaced by the
noncommutative ones 
\begin{equation}
\lbrack x_{\mu }\,,x_{\nu }]=0\qquad \Longrightarrow \qquad \lbrack \hat{x}%
_{\mu }\,,\hat{x}_{\nu }]=\frac{i}{\kappa ^{2}}\Theta _{\mu \nu }(\kappa 
\hat{x})\,,  \label{nnn}
\end{equation}%
where $\Theta _{\mu \nu }$ is a given tensorial function\footnote{%
We add that one can also consider interesting models with the function $%
\Theta _{\mu \nu }$ treated as new dynamical "noncommutativity" field (see
e.g. \cite{arr}).}%
\begin{equation}
\Theta _{\mu \nu }(\kappa \hat{x})=\Theta _{\mu \nu }^{(0)}+\kappa \Theta
_{\mu \nu }^{(1)\rho }\widehat{x}_{\rho }+\cdots ,
\end{equation}%
and $\kappa $ introduced as geometric mass-like parameter. The simplest
choice $\Theta _{\mu \nu }(\kappa \widehat{x})=\Theta _{\mu \nu }^{(0)}$ \
corresponds to the canonical (DFR\footnote{%
DFR$\equiv $Dopplicher-Fredenhagen-Roberts} \cite{lit1} or Moyal-Weyl, see
e.g. \cite{dlw6}-\cite{bala}) noncommutative space-time and $\Theta _{\mu
\nu }(\kappa \widehat{x})=\Theta _{\mu \nu }^{(1)\rho }\widehat{x}_{\rho }$
describes quantum space-time with Lie-algebraic noncommutativity (e.g. $%
\kappa $-deformed Minkowski \cite{zak}-\cite{7}).

Firstly NC free fields were introduced with keeping the classical Poincare
symmetry and standard field oscillators algebra unchanged - such approach we
call "traditional". The new approach started later, with the introduction of
deformed quantum Poincare symmetries as determining the formulation of NC
field theory, with quantum covariance implying definite modification of
field oscillators algebra.

1.\underline{\textit{Traditional approach to NC QFT}}. \ Such formalism was
firstly elaborated in \cite{lit1}, \cite{fil} for the canonical
noncommutative space-time. In such approach it is postulated that the
classical Poincare symmetries remain undeformed, with the noncommutativity (%
\ref{nnn}) introducing the breaking of classical Lorentz invariance. The NC
quantum free fields $\widehat{\varphi }(\widehat{x})$\footnote{%
We denote by $\widehat{\varphi }(x)$ and $\widehat{\varphi }(\widehat{x})$
respectively quantum free fields on standard and noncommutative Minkowski
spaces in traditional approach. The standard classical free fields are
denoted by $\varphi (x)$; if we substitute $x_{\mu }\rightarrow \widehat{x}%
_{\mu }$ we obtain the classical NC free fields denoted by $\varphi (%
\widehat{x}).$} are obtained by the replacement $x_{\mu }\rightarrow 
\widehat{x}_{\mu }$ (see (\ref{nnn})) inserted in the standard quantum free
scalar KG field $\widehat{\varphi }(x)$\footnote{%
In this paper for simplicity we shall consider only scalar fields. The
standard creation and annihilation operators are defined respectively by $a(%
\overrightarrow{p},p_{0}=\omega (\overrightarrow{p}))$ and $a(%
\overrightarrow{p},p_{0}=-\omega (\overrightarrow{p}))$, where $\omega (%
\overrightarrow{p})=\sqrt{\overrightarrow{p}^{2}+m^{2}}$ and for real fields
(\ref{fffd}) $a(\overrightarrow{p},p_{0})$ = $a^{\dagger }(-\overrightarrow{p%
},-p_{0})$.}%
\begin{equation}
\widehat{\varphi }({x})=\frac{1}{(2\pi )^{4}}\int d^{4}p\delta
(p^{2}-m^{2})\;\mathrm{e}^{ip{x}}a(p),  \label{fffd}
\end{equation}%
where the quantized field oscillators satisfy the undeformed (standard)
covariant oscillator algebra $\widehat{\mathcal{H}}_{0}$ with the following
binary commutation relations%
\begin{equation}
\delta (p^{2}-m^{2})\delta (q^{2}-m^{2})[a(p),a(q)]=\epsilon (p_{0})\delta
(p^{2}-m^{2})\delta ^{(4)}(p+q).  \label{sssa}
\end{equation}%
\ One can use the Weyl map representing the algebra $\widehat{\mathcal{M}}$
of functions $\widehat{f}\equiv f(\widehat{x})$ on noncommutative space-time
(see e.g. \cite{blo},\cite{pietka})%
\begin{equation}
\widehat{\mathcal{M}}(f(\widehat{x}),\cdot )\overset{W}{\simeq }\mathcal{M}%
(f(x),\star ),  \label{wm}
\end{equation}%
with suitable nonlocal $\star -$multiplication. The NC field theory due to
the homomorphic Weyl mapping can be represented as a nonlocal theory of
standard fields on Minkowski space-time. After the extension of Weyl map (%
\ref{wm}) to the products of functions depending on different copies of
noncommutative Minkowski spaces one can map the algebra of NC quantum fields
into the multilocal algebra of standard quantum fields, in particular%
\footnote{%
In (\ref{www}) we describe only the Weyl map for binary field products.} 
\begin{equation}
\widehat{\varphi }(\widehat{x})\widehat{\varphi }(\widehat{y})\overset{W}{%
\simeq }\widehat{\varphi }(x)\star \widehat{\varphi }(y).  \label{www}
\end{equation}

Using (\ref{www}) one can map the commutator of NC quantum field into the $%
\star -$commutator of standard quantum fields described in usual Minkowski
space by the formula $([A,B]_{\star }:=A\star B-B\star A)$%
\begin{equation}
\lbrack \widehat{\varphi }(\widehat{x}),\widehat{\varphi }(\widehat{y})]%
\overset{W}{\simeq }[\widehat{\varphi }(x),\widehat{\varphi }(y)]_{\star
}=i\Delta _{\star }(x,y;m^{2}),
\end{equation}%
where 
\begin{eqnarray}
\Delta _{\star }(x,y;m^{2}) &=&\frac{-i}{(2\pi )^{3}}\int d^{4}p\delta
(p^{2}-m^{2})\delta (q^{2}-m^{2})[a(p)a(q)\mathrm{e}^{ip{x}}\,\star \mathrm{e%
}^{iq{y}}  \label{npj} \\
&&\qquad \qquad \qquad \qquad \qquad \ \ \ \ \ \ \ \ \ \ \ \ \ \ \ \ \ \ \ \
\ \ \ \ \ -a(q)a(p)\mathrm{e}^{iq{y}}\,\star \mathrm{e}^{ipx}].  \notag
\end{eqnarray}%
Because $\star $-product $\mathrm{e}^{ip{x}}\,\star \mathrm{e}^{ip{y}}$ even
in simplest case of canonical noncommutativity (\ref{nnn}) is not symmetric,
the commutator (\ref{sssa}) can not be factored out and the $\star $%
-commutator (\ref{npj})\ is not a $c$-number. In canonical (Moyal-Weyl) case
however one can obtain from (\ref{npj}) the Pauli-Jordan relativistic
commutator function%
\begin{equation}
\Delta (x-y;m^{2})=\frac{i}{(2\pi )^{4}}\int d^{4}p\epsilon (p_{0})\delta
(p^{2}-m^{2})\mathrm{e}^{ip{x}},  \label{spjs}
\end{equation}%
if we suitably modify the standard oscillator algebra (\ref{sssa}) (see \cite%
{wess}-\cite{bala}).

In traditional NC QFT the standard relativistic locality or microcausality
condition can be replaced by the vanishing of $\star -$commutator for
space-like separations of points $x$ and $y$ $((x-y)^{2}<0)$ (see e.g. \cite%
{wess},\cite{pietka}), defining $\star $-locality%
\begin{equation}
\lbrack \widehat{\varphi }(x),\widehat{\varphi }(y)]=0\overset{NC}{%
\longrightarrow }[\widehat{\varphi }(x),\widehat{\varphi }(y)]_{\star }=0.
\label{fd}
\end{equation}

The traditional approach to NC QFT was further formalized as a modification
of Wightmann framework of QFT, with the interpretation of modified locality (%
\ref{fd}) in terms of so-called wedged geometries \ (see e.g. \cite{gl}).

2. $\underline{\text{\textit{\text{New approach to NC QFT with quantum
Poincare symmetries}}}}.$ Such approach firstly appeared with the
introduction of canonical noncommutativity relations $(\Theta _{\mu \nu
}(\kappa \widehat{x})=\Theta _{\mu \nu }^{(0)}$ in (\ref{nnn})$)$ as
generated by twist factor%
\begin{equation}
\mathcal{F}=e^{\frac{\,i}{2}\,\theta ^{\mu \nu }\,P_{\mu }\otimes P_{\nu
}}\,\,,  \label{twist0}
\end{equation}%
which determines uniquely the corresponding Hopf-algebraic canonical
deformation of Poincare symmetries \cite{dlw6}-\cite{wess},\cite{pietka}. In
the framework of twisted Poincare symmetries the wide class of relations (%
\ref{nnn}) are covariant under the action of suitably chosen quantum
Poincare algebra generators (see e.g. \cite{lw}). If the twist $\mathcal{F=F}%
_{(1)}\otimes \mathcal{F}_{(2)}$ is known the $\star -$product of NC quantum
fields in new approach is determined by the formulae \cite{dlw6}-\cite{wess} 
\begin{equation}
\widehat{\phi }(\widehat{x})\widehat{\phi }(\widehat{y})\overset{W}{\simeq }%
\widehat{\phi }(x)\star \widehat{\phi }(y)\equiv m(\mathcal{F}^{-1}\circ
\lbrack \widehat{\phi }(x)\otimes \widehat{\phi }(y)]),  \label{ccdf}
\end{equation}%
where we denote by $\widehat{\phi }(x)$ and $\widehat{\phi }(\widehat{x})$
the quantum fields respectively on commutative and noncommutative
space-time, which transform under deformed (quantum) Poincare symmetries.

In new approach to NC QFT the algebra of NC fields should be
quantum-covariant, i.e. covariant under the Hopf-algebraic action of
generators describing quantum relativistic symmetries. In particular it
follows that in order to introduce quantum-deformed covariant free field
quantization, we should replace the standard field commutator by its
quantum-deformed braided version \cite{zahn},\cite{alv}. Further, following
several authors (\cite{wess},\cite{pietka}, \cite{11c}-\cite{fi2}), in
quantum-covariant formulation of NC quantum fields with quantum Poincare
symmetries characterized by the universal $\mathcal{R}$-matrix $\mathcal{R}=%
\mathcal{R}_{(1)}\otimes \mathcal{R}_{(2)}$, we shall replace the standard
commutators in (\ref{sssa}) by the following braided commutators defining
the field oscillators algebra $\widehat{\mathcal{H}}$%
\begin{equation}
\lbrack a(p),a(q)]\qquad \longrightarrow \qquad \lbrack
A(p),A(q)]^{BR}\equiv A(p)A(q)-(\mathcal{R}_{(2)}\blacktriangleright A(q))(%
\mathcal{R}_{(1)}\blacktriangleright A(p)),  \label{bra}
\end{equation}%
where $\blacktriangleright $ describes the action on the module $\widehat{%
\mathcal{H}}$ of the deformed Poincare generators.

The aim of this paper is to study the NC quantum fields which are covariant
under the quantum-deformed Poincare symmetries, in particular the
determination of new algebraic structure of the algebra of such NC quantum
free fields $\Phi (\widehat{\phi }(\widehat{x}),\bullet )$ with suitably
deformed new $\bullet $-multiplication. For the products of functions on
noncommutative Minkowski space we will employ the Weyl map (\ref{wm}) with
its multilocal extension and express the NC quantum fields as nonlocal QFT
on classical Minkowski space. Following main ideas of our recent paper \cite%
{lwnaj} we shall present the formulation of quantum covariant free NC
quantum fields with the introduction of necessary braidings.

In order to introduce the quantum-covariant theory of NC fields one should
use the general covariance properties of tensor product $U\otimes V$ of pair
of quantum Poincare algebra modules $U$ and $V$. If the quantum Poincare
algebra is characterized by universal matrix $\mathcal{R}=\mathcal{R}%
_{(1)}\otimes \mathcal{R}_{(2)}$, the transposed tensor product of its
modules is given by the following braided transposition\footnote{%
Below in Sect. 2 we shall distinguish two different actions of Poincare
algebra generators. In formulas (\ref{uv}) and (\ref{dfd}) the action $%
\vartriangleright $ is still not specified.} \ (see e.g. \cite{sm})%
\begin{equation}
\Psi (U\otimes V)=(\mathcal{R}_{(2)}\rhd V)\otimes (\mathcal{R}_{(1)}\rhd U),
\label{uv}
\end{equation}%
where $\Psi $ is the intertwiner of quantum-deformed Poincare algebra
modules. If we choose $U=\widehat{\phi }(\widehat{x}),$ $V=\widehat{\phi }(%
\widehat{y})$ the quantum-deformed covariant commutator takes the braided
form%
\begin{equation}
\lbrack \widehat{\phi }(\widehat{x}),\widehat{\phi }(\widehat{y})]_{\bullet
}^{BR}\equiv \widehat{\phi }(\widehat{x})\bullet \widehat{\phi }(\widehat{y}%
)-(\mathcal{R}_{(2)}\rhd \widehat{\phi }(\widehat{y}))\bullet (\mathcal{R}%
_{(1)}\rhd \widehat{\phi }(\widehat{x})).  \label{dfd}
\end{equation}%
where the action $\rhd $ on NC fields will be specified later in Sect. 2.
The form (\ref{dfd}) of braided field commutator with local standard
multiplication was proposed firstly by Oeckl \cite{dlw6}, further used by
Zahn \cite{zahn} and advocated \ by Aschieri at all \cite{alv}; one should
comment however that the relation (\ref{dfd}) is not in common use in NC QFT.

Further we shall use the formula for universal $\mathcal{R}$-matrix
describing quantum twist-deformed Poincare symmetries, given by the relations%
\begin{equation}
\mathcal{R}=\mathcal{F}_{21}\mathcal{F}^{-1}=\mathcal{F}^{-2},\qquad 
\mathcal{F}_{21}=\mathcal{F}_{(2)}\otimes \mathcal{F}_{(1)}.  \label{rfr}
\end{equation}

In order to discuss the braided structure of NC quantum\ field theory
covariant under quantum Poincare symmetries we shall introduce in Sect. 2
the actions of quantum Poincare algebra generators on the algebra $\Phi (%
\widehat{{\phi }}(\widehat{x}),\bullet )$ of quantum-deformed NC fields. We
shall define braided covariant product $\widehat{\phi }(\widehat{x})\bullet 
\widehat{\phi }(\widehat{y})$ with braid factor (\ref{uv}) describing the
exchange of oscillators and noncommutative Fourier exponentials. In Sect.3
for the case of Moyal-Weyl space-time noncommutativity we shall consider in
detail the braided quantization of free scalar quantum NC fields and discuss
the braided locality. In last Section in particular we point out that there
is alternative way \cite{fi},\cite{fi2} of defining the braiding in the
noncommutative quantum field commutators (\ref{dfd}) which leads to the
triviality of braid factor.

\section{\protect\bigskip The covariance of the algebra of quantum NC fields}

In the description of the algebra of quantum NC fields one should answer the
following two questions:

\begin{itemize}
\item how to define single quantum NC field as describing a representation
(module) of quantum Poincare algebra

\item how to define the products of quantum NC fields in covariant way, i.e.
as a tensorial representation (module) of quantum Poincare algebra (the
answer should be given at least for the binary products)
\end{itemize}

\subsection{Single quantum NC fields as quantum Poincare algebra module}

The quantum NC field $\widehat{\phi }(\widehat{x})$ can be described as
infinite sum (in fact continuous integral) of tensor products of
noncommutative plane waves $e^{ip\widehat{x}}$ describing the basis of
algebra $\widehat{\mathcal{M}}$ and the elements $a(p)$ determining the
algebra of field oscillators $\widehat{\mathcal{H}}$%
\begin{equation}
\widehat{{\phi }}(\widehat{x})\in \widehat{\mathcal{M}}\underline{\otimes }%
\widehat{\mathcal{H}},  \label{fie}
\end{equation}%
where $\underline{\otimes }$ denotes braided tensor product with braided $%
\bullet -$multiplication (see Sect. 2.2). Using the homomorphic Weyl map of
the algebra $\widehat{\mathcal{M}}$ (see (\ref{wm})) one can represent the
noncommutative algebra $\widehat{\mathcal{M}}$ by the algebra $\mathcal{M}$
of classical functions with $\star -$multiplication law. The Weyl map can be
applied to the first factor $\widehat{\mathcal{M}}$ in (\ref{fie}) and leads
to the $\widehat{\mathcal{H}}$-algebra-valued representation of NC quantum
fields, i.e.%
\begin{equation}
\widehat{{\phi }}{(}\widehat{x})\overset{W}{\simeq }\widehat{{\phi }}{(}%
x)\in \widehat{\mathcal{H}},  \label{wew}
\end{equation}%
because after the replacement in the Fourier expansion of $\widehat{{\phi }}(%
\widehat{x})$ the basis $e^{ip\widehat{x}}$ by $e^{ipx}$ one can use the
isomorphism $1\underline{\otimes }\widehat{\mathcal{H}}\simeq 1\otimes 
\widehat{\mathcal{H}}\simeq \widehat{\mathcal{H}}$. It appears that the
algebra of field operators $\widehat{{\phi }}(x)$ have well defined
no-deformation limit ($\widehat{{\phi }}{(}x)\rightarrow \widehat{\varphi }%
(x)$), with the algebra $\widehat{\mathcal{H}}$ becoming the algebra $%
\widehat{\mathcal{H}}_{0}$ (see (\ref{sssa})).

\bigskip Let us specify now the action of the deformed Poincare algebra
generators on single NC\ quantum field (\ref{fie}). We shall recall firstly
two possible actions of classical Poincare algebra on standard free quantum
fields $\widehat{{\varphi }}(x)\in \widehat{\mathcal{H}}_{0}$ (see (\ref%
{fffd}),(\ref{sssa})).

\begin{enumerate}
\item classical differential space-time realization on the functions on
classical Minkowski space-time%
\begin{equation}
P_{\mu }\vartriangleright \widehat{{\varphi }}(x)=\frac{1}{i}\partial _{\mu }%
\widehat{{\varphi }}(x),\qquad M_{\mu \nu }\vartriangleright \widehat{{%
\varphi }}(x)=\frac{1}{i}x_{[\mu }\partial _{\nu ]}\widehat{{\varphi }}(x),
\label{rd1}
\end{equation}

\item quantum-mechanical realization on the free field oscillators algebra%
\begin{equation}
P_{\mu }\blacktriangleright \widehat{{\varphi }}(x)=[P_{\mu },\widehat{{%
\varphi }}(x)],\qquad M_{\mu \nu }\blacktriangleright {\phi }=[M_{\mu \nu },%
\widehat{{\varphi }}(x)],  \label{rd2}
\end{equation}

with $P_{\mu },M_{\mu \nu }\in \widehat{\mathcal{H}}_{0}$.
\end{enumerate}

The classical Poincare covariance relation%
\begin{equation}
U(\Lambda ,a)\widehat{{\varphi }}(x)U^{-1}(\Lambda ,a)=\widehat{{\varphi }}%
(\Lambda x+a),
\end{equation}%
where $U(\Lambda ,a)=\exp (ia^{\mu }P_{\mu }+i\omega ^{\mu \nu }M_{\mu \nu
}) $ for infinitesimal $a_{\mu }$ and $\omega _{\mu \nu }$ ($\Lambda _{\ \nu
}^{\mu }=\delta _{\ \nu }^{\mu }+\omega _{\ \nu }^{\mu }$) links two
realizations (\ref{rd1}),(\ref{rd2}). It leads to the following
infinitesimal covariance condition%
\begin{equation}
g\rhd \widehat{{\varphi }}(x)=-g\blacktriangleright \widehat{{\varphi }}%
(x),\qquad g=(P_{\mu },M_{\mu \nu }).  \label{ccd}
\end{equation}

We see therefore that for Poincare-covariant standard (undeformed) quantum
fields one can use as the action of classical Poincare algebra generators
equivalently the "classical" action $\rhd $ or the quantum-mechanical one $%
\blacktriangleright $.

Now we pass to quantum-deformed NC fields (\ref{fie}). At final stage of
considerations {we shall consider such fields after the Weyl map (\ref{wew})}%
. The realizations (\ref{rd1}) and (\ref{rd2}) due to deformation are
modified, however in covariant theory the covariance condition (\ref{ccd})
remains valid provided that we modify the relation (\ref{ccd}) as follows ($%
S $ is an antipode)%
\begin{equation}
g\rhd \widehat{{\phi }}(\widehat{x})=S(g)\blacktriangleright \widehat{{\phi }%
}(\widehat{x}).  \label{sss}
\end{equation}%
By analogy with undeformed case in the tensor product $\widehat{f}\otimes 
\widehat{h}\in \widehat{{\phi }}(\widehat{x})$ ($\widehat{f}\in \widehat{%
\mathcal{M}},$ $\widehat{h}\in \widehat{\mathcal{H}}$) the actions $%
\vartriangleright $ on $\widehat{h}$ and $\blacktriangleright $ on $\widehat{%
f}$ are assumed to be trivial:%
\begin{equation}
g\rhd \widehat{{h}}=\epsilon (g)\widehat{{h}},\qquad g\blacktriangleright 
\widehat{{f}}=\epsilon (g)\widehat{{f}}.  \label{28}
\end{equation}%
If we use Hopf-algebraic formula (the case of action $\blacktriangleright $
is analogous)%
\begin{equation}
g\rhd (\widehat{f}\otimes \widehat{h})=\Delta (g)\rhd (\widehat{f}\otimes 
\widehat{h})=(g_{(1)}\rhd \widehat{f})\otimes (g_{(2)}\rhd \widehat{h})\,\,,
\end{equation}%
the actions $g\rhd \widehat{{\phi }}(\widehat{x})$ and $g\blacktriangleright 
\widehat{{\phi }}(\widehat{x})$ in (\ref{sss}) due to the relations (\ref{28}%
) and the structure of coproduct $\Delta (g)$ with unique terms $g\otimes 1$%
and $1\otimes g$ take the form%
\begin{equation}
g\rhd (\widehat{f}\otimes \widehat{h})\equiv (g\rhd \widehat{f})\otimes 
\widehat{h},\qquad \qquad g\blacktriangleright (\widehat{f}\otimes \widehat{h%
})\equiv \widehat{f}\otimes (g\blacktriangleright \widehat{h}).  \label{27}
\end{equation}

After applying the Weyl map (\ref{wew}) one can rewrite (\ref{sss}) in the
form similar to (\ref{ccd}) 
\begin{equation}
g\rhd \widehat{{\phi }}(x)=S(g)\blacktriangleright \widehat{{\phi }}(x),
\label{uiu}
\end{equation}%
where commutators in (\ref{rd2}) should be replaced by quantum adjoint action%
\begin{equation}
g\blacktriangleright \widehat{{\varphi }}(x)=[g,\widehat{\varphi }(x)]%
\overset{\text{quantum}}{\underset{\text{deformation}}{\Longrightarrow }}%
g\blacktriangleright \widehat{{\phi }}(\widehat{x})=ad_{g}\widehat{{\phi }}(%
\widehat{x})=g_{(1)}\widehat{{\phi }}(\widehat{x})S(g_{(2)}).  \label{adds}
\end{equation}

If the Hopf-algebraic form of deformed Poincare algebra is known, the
formula (\ref{adds}) is fully determined (S denotes the antipode);
subsequently the action $g\rhd \widehat{{\phi }}(x)$ described by the
deformation of (\ref{rd1}) should be chosen in consistency with the relation
(\ref{uiu}).

It should be added that one can introduce third possible action $%
\trianglerighteq $ of generators on NC quantum fields (see e.g. \cite{fi}, 
\cite{fi2}), defined by the formula%
\begin{equation}
g\trianglerighteq (\widehat{f}\otimes \widehat{h})=(g_{(1)}\rhd \widehat{f}%
)\otimes (g_{(2)}\blacktriangleright \widehat{h})\,\,.  \label{nn1}
\end{equation}%
Such action of generators if applied to the field $\widehat{{\phi }}(%
\widehat{x})$ due to the relations (\ref{27}) leads to the following form of
covariance conditions (\ref{sss})%
\begin{equation}
g\trianglerighteq \widehat{{\phi }}(\widehat{x})=\epsilon (g)\widehat{{\phi }%
}(\widehat{x})=0,  \label{nn2}
\end{equation}%
which after the Weyl map (\ref{wew}) provides the covariance relation (\ref%
{uiu}).

\subsection{The covariant action of deformed Poincare algebra on the product
of NC quantum fields}

In order to formulate the deformed NC QFT we shall define firstly the
algebra $\Phi (\widehat{\phi }(\widehat{x}),\bullet )$ of NC fields $%
\widehat{{\phi }}(\widehat{x})$ and further perform the Weyl map (see (\ref%
{wew})). We multiply the NC fields (\ref{fie}) using the new braided $%
\bullet -$multiplication which defines the multiplication of NC quantum
fields with deformed field oscillators algebra%
\begin{eqnarray}
\widehat{{\phi }}(\widehat{x})\bullet \widehat{{\phi }}(\widehat{y}) &=&m_{%
\mathcal{M}\underline{\otimes }\mathcal{H}}\widehat{{\phi }}(\widehat{x})%
\underline{\otimes }\widehat{{\phi }}(\widehat{y})  \label{ggn} \\
&=&(m_{\mathcal{M}}\otimes m_{\mathcal{H}})\circ (id\otimes \Psi _{\mathcal{M%
},\mathcal{H}}\otimes id)[\widehat{{\phi }}(\widehat{x})\otimes \widehat{{%
\phi }}(\widehat{y})].  \notag
\end{eqnarray}%
Braid factor $\Psi _{\mathcal{M},\mathcal{H}}$ describes effectively the
noncommutativity of factors $A(p)$ and $e^{iq\widehat{y}}$ in the product of
field operators $\widehat{{\phi }}(\widehat{x})$ and $\widehat{{\phi }}(%
\widehat{y})$ and it is needed in general case in order to obtain the
product (\ref{ggn}) as an element of $\widehat{\mathcal{M}}^{(2)}\underline{%
\otimes }\widehat{\mathcal{H}}$ where by $\widehat{\mathcal{M}}^{(n)}$ we
denote the noncommutative functions on the $n$-tuple of quantum Minkowski
spaces ($\widehat{x}^{(1)},\widehat{x}^{(2)},\ldots ,\widehat{x}^{(n)}$)
(for $n=2$ we have chosen $\widehat{x}^{(1)}=\widehat{x},$ $\widehat{x}%
^{(2)}=\widehat{y}$). The formula (\ref{ggn}) permits to express the basis
of binary field products (\ref{ggn})%
\begin{equation}
A=m_{\mathcal{M}\underline{\otimes }\mathcal{H}}[(e^{ip\widehat{x}}A(p))%
\underline{\otimes }(e^{iq\widehat{y}}A(q))]=(e^{ip\widehat{x}}A(p))\bullet
(e^{iq\widehat{y}}A(q)),  \label{3a}
\end{equation}%
by the superposition of elements of tensor product $\widehat{\mathcal{M}}%
^{(2)}\otimes \widehat{\mathcal{H}}$ spanned by the elements%
\begin{equation}
\widetilde{A}=e^{ip\widehat{x}}e^{iq\widehat{y}}\underline{\otimes }%
A(p)A(q)\in \widehat{\mathcal{M}}^{(2)}\otimes \widehat{\mathcal{H}}.
\label{3b}
\end{equation}%
Further we assume that the quantum deformation of Poincare algebra is
described by quasi-triangular Hopf algebra characterized by the universal $%
\mathcal{R}$-matrix $\mathcal{R}=\mathcal{R}_{(1)}\otimes \mathcal{R}_{(2)}.$
Following the general formula (\ref{uv}) we introduce the braid factor
expressing the transposition of noncommutative plane waves and deformed
field oscillators%
\begin{equation}
\Psi _{\mathcal{M},\mathcal{H}}[A(p)\underline{\otimes }e^{iq\widehat{y}}]=(%
\mathcal{R}_{(2)}\rhd e^{iq\widehat{y}})\underline{\otimes }(\mathcal{R}%
_{(1)}\blacktriangleright A(p)).  \label{fun}
\end{equation}%
Using (\ref{ggn}) leads to the following equivalent expression for the
product (\ref{3a})%
\begin{equation}
A=[e^{ip\widehat{x}}\mathcal{R}_{(2)}\rhd e^{iq\widehat{y}}]\underline{%
\otimes }[(\mathcal{R}_{(1)}\blacktriangleright A(p))A(q)].  \label{jjj}
\end{equation}

We recall that we have used here universality of the formula (\ref{uv}) for
any two deformed Poincare algebra modules, i.e. the general formula (\ref%
{ggn}) can be rewritten in concrete way as follows 
\begin{equation}
(\widehat{f}\otimes \widehat{h})\bullet (\widehat{f}^{\prime }\otimes 
\widehat{h}^{\prime })=[\widehat{f}\cdot \mathcal{R}_{(2)}\rhd \widehat{f}%
^{\prime }]\underline{\otimes }[(\mathcal{R}_{(1)}\blacktriangleright 
\widehat{h})\cdot \widehat{h}^{\prime }].  \label{bmm}
\end{equation}%
where $\widehat{f},\widehat{f}^{\prime }\in \widehat{\mathcal{M}}$ and $%
\widehat{h},\widehat{h}^{\prime }\in \widehat{\mathcal{H}}$. If we perform
the Weyl map in the algebra $\widehat{\mathcal{M}}^{(2)}$ ($\widehat{f}%
\overset{W}{\rightarrow }f,\widehat{f}^{\prime }\overset{W}{\rightarrow }%
f^{\prime }$) and introduce corresponding star product the first factor in
the tensor product on rhs of (\ref{bmm}) is becoming a classical function in
accordance with the prescription\footnote{%
For notational convenience one can introduce the symbol $\circledast $ by
means of the formula $(\widehat{f}\otimes \widehat{h})\bullet (\widehat{f}%
^{\prime }\otimes \widehat{h}^{\prime })\overset{W}{\simeq }(f\otimes 
\widehat{h})\circledast (f^{\prime }\otimes \widehat{h}^{\prime })$. The
translation of algebraic properties of $\bullet -$multiplication (\ref{ggn})
(e.g. associativity) into the corresponding properties of $\circledast $ is
under consideration.}%
\begin{equation}
\widehat{f}\cdot (\mathcal{R}_{(2)}\rhd \widehat{f}^{\prime })\overset{W}{%
\simeq }f\star (\mathcal{R}_{(2)}\rhd f^{\prime }),
\end{equation}%
and we obtain that%
\begin{equation}
(\widehat{f}\otimes \widehat{h})\bullet (\widehat{f}^{\prime }\otimes 
\widehat{h}^{\prime })\overset{W}{\simeq }(f\star (\mathcal{R}_{(2)}\rhd
f^{\prime })[(\mathcal{R}_{(1)}\blacktriangleright \widehat{h})\cdot 
\widehat{h}^{\prime }],  \label{hhf}
\end{equation}%
if the relation $%
\mathbb{C}
\underline{\otimes }\widehat{\mathcal{H}}\simeq \widehat{\mathcal{H}}$ is
used.

As we mentioned earlier, we shall use the covariant braided field
commutator\ (\ref{dfd}), which after using specified actions $%
\vartriangleright ,\blacktriangleright $ can be rewritten more explicitly%
\begin{equation}
\lbrack \widehat{\phi }(\widehat{x}),\widehat{\phi }(\widehat{y})]_{\bullet
}^{BR}=\widehat{\phi }(\widehat{x})\bullet \widehat{\phi }(\widehat{y})-[(%
\mathcal{R}_{(2)}\otimes 1)\triangleright \widehat{\phi }(\widehat{y}%
)]\bullet \lbrack (1\otimes \mathcal{R}_{(1)})\blacktriangleright \widehat{%
\phi }(\widehat{x})].  \label{big}
\end{equation}

The formula (\ref{big}) is our basic relation which defines the commutator
of free NC quantum fields. We shall show below that the quantum covariance
of the product (\ref{ggn}) and of deformed commutator requires braid (\ref%
{fun}) and braided commutator (\ref{hhf})). For covariant NC quantum fields
it follows however from (\ref{sss}) 
\begin{equation}
(\mathcal{R}_{(2)}\otimes 1)\triangleright \widehat{\phi }(\widehat{y}%
)=(1\otimes S(\mathcal{R}_{(2)}))\blacktriangleright \widehat{\phi }(%
\widehat{y}),  \label{41}
\end{equation}%
i.e. it follows that braided field commutator (\ref{big}) can be replaced by
other two ways which employs only the action $\vartriangleright $ or $%
\blacktriangleright $\footnote{%
We use shorthand notation $\mathcal{R}_{21}\triangleright (a\bullet b)=(%
\mathcal{R}_{(2)}\triangleright a)\bullet (\mathcal{R}_{(1)}\triangleright b)
$ etc.}%
\begin{eqnarray}
\lbrack \widehat{\phi }(\widehat{x}),\widehat{\phi }(\widehat{y})]_{\bullet
}^{BR} &=&\widehat{\phi }(\widehat{x})\bullet \widehat{\phi }(\widehat{y})-%
\mathcal{R}_{21}\triangleright (\widehat{\phi }(\widehat{y})\bullet \widehat{%
\phi }(\widehat{x}))  \label{cdc} \\
\lbrack \widehat{\phi }(\widehat{x}),\widehat{\phi }(\widehat{y})]_{\bullet
}^{BR} &=&\widehat{\phi }(\widehat{x})\bullet \widehat{\phi }(\widehat{y})-%
\mathcal{R}_{21}\blacktriangleright (\widehat{\phi }(\widehat{y})\bullet 
\widehat{\phi }(\widehat{x})).  \label{cdc1}
\end{eqnarray}%
Further we shall employ formula (\ref{41}) with the quantum-mechanical
action $\blacktriangleright $. Let us demonstrate firstly that the field
product (\ref{ggn}) is covariant under the action of a quantum Poincare
generator $g\in \mathcal{U}(\mathcal{P}_{4})$. Using the Hopf-algebraic
formula with fourfold coproduct $\Delta ^{(4)}(g)=g_{(1)}\otimes
g_{(2)}\otimes g_{(3)}\otimes g_{(4)}$ and the identity $%
g\blacktriangleright (a\blacktriangleright b)=(ga)\blacktriangleright b$ we
get%
\begin{eqnarray}
g\blacktriangleright \lbrack (\widehat{f}\otimes \widehat{h})\bullet (%
\widehat{f}^{\prime }\otimes \widehat{h}^{\prime })]
&=&[(g_{(1)}\blacktriangleright \widehat{f})\cdot (g_{(2)}\mathcal{R}%
_{(2)}\blacktriangleright \widehat{f}^{\prime })]\otimes \lbrack (g_{(3)}%
\mathcal{R}_{(1)}\blacktriangleright \widehat{h})\cdot
(g_{(4)}\blacktriangleright \widehat{h}^{\prime })]  \notag \\
&=&[(g_{(1)}\blacktriangleright \widehat{f})\cdot (\mathcal{R}%
_{(2)}g_{(3)}\blacktriangleright \widehat{f}^{\prime })]\otimes \lbrack (%
\mathcal{R}_{(1)}g_{(2)}\blacktriangleright \widehat{h})\cdot
(g_{(4)}\blacktriangleright \widehat{h}^{\prime })]  \notag \\
&=&[(g_{(1)}\blacktriangleright \widehat{f})\cdot
(g_{(2)}\blacktriangleright \widehat{h})]\bullet \lbrack
(g_{(3)}\blacktriangleright \widehat{f}^{\prime })\cdot
(g_{(4)}\blacktriangleright \widehat{h}^{\prime })]  \notag \\
&=&[g_{(1)}\blacktriangleright (\widehat{f}\cdot \widehat{h})]\bullet
\lbrack g_{(2)}\blacktriangleright (\widehat{f}^{\prime }\cdot \widehat{h}%
^{\prime })]  \label{pppr}
\end{eqnarray}%
where we use the relations (\ref{28}) and the equalities\footnote{%
See e.g. \cite{maj}, Sec . 9.2. We use Sweedler notation with suppressed
summation indices. If more explicitly $\Delta
^{(4)}(g)=\sum_{I}g_{(1)}^{I}\otimes g_{(2)}^{I}\otimes g_{(3)}^{I}\otimes
g_{(4)}^{I}$ and $\mathcal{R}^{(4)}=\sum_{J}1\otimes \mathcal{R}%
_{(1)}^{J}\otimes \mathcal{R}_{(2)}^{J}\otimes 1$ we read e.g. eq. (\ref{iji}%
) as $\sum_{I}\sum_{J}g_{(2)}^{I}\mathcal{R}_{(2)}^{J}\otimes g_{(3)}^{I}%
\mathcal{R}_{(1)}^{J}=\sum_{I}\sum_{J}\mathcal{R}_{(2)}^{J}g_{(3)}^{I}%
\otimes \mathcal{R}_{(1)}^{J}g_{(2)}^{I}.$}.%
\begin{equation}
g_{(2)}^{I}\mathcal{R}_{(2)}^{J}\otimes g_{(3)}^{I}\mathcal{R}_{(1)}^{J}=%
\mathcal{R}_{(2)}^{J}g_{(3)}^{I}\otimes \mathcal{R}_{(1)}^{J}g_{(2)}^{I}.
\label{iji}
\end{equation}%
Subsequently, applying (\ref{pppr}) to the basis (\ref{3a}) of the product
of two NC quantum fields one gets its quantum-Poincare covariance%
\begin{equation}
g\blacktriangleright (\widehat{\phi }(\widehat{{x}})\bullet \widehat{\phi }(%
\widehat{{y}}))=m_{\mathcal{\bullet }}[\Delta (g)\blacktriangleright 
\widehat{\phi }(\widehat{x})\otimes \widehat{\phi }(\widehat{{y}}%
)]=(g_{(1)}\blacktriangleright \widehat{\phi }(\widehat{x}))\bullet
(g_{(2)}\blacktriangleright \widehat{\phi }(\widehat{{y}})).  \label{f1}
\end{equation}%
Now we shall show the covariance of the braided field commutator. The action
of generator $g$ on second term defining braided commutator (\ref{big}) with
action $\blacktriangleright $ is%
\begin{align}
g\blacktriangleright (\mathcal{R}_{21}\blacktriangleright \lbrack \widehat{%
\phi }(\widehat{{y}})\bullet \widehat{\phi }(\widehat{x})])&
=g\blacktriangleright \lbrack (\mathcal{R}_{(2)}\blacktriangleright \widehat{%
\phi }(\widehat{{y}}))\bullet (\mathcal{R}_{(1)}\blacktriangleright \widehat{%
\phi }(\widehat{x}))]  \label{f2} \\
& =m_{\mathcal{\bullet }}(\Delta (g)\mathcal{R}_{21}\blacktriangleright 
\widehat{\phi }(\widehat{{y}})\otimes \widehat{\phi }(\widehat{x})).  \notag
\end{align}%
The covariance of braided commutator (\ref{big}) requires that%
\begin{equation}
g\blacktriangleright (\mathcal{R}_{21}\blacktriangleright (\widehat{\phi }(%
\widehat{y})\bullet \widehat{\phi }(\widehat{x})))=\mathcal{R}%
_{21}\blacktriangleright (g\blacktriangleright (\widehat{\phi }(\widehat{y}%
)\bullet \widehat{\phi }(\widehat{x}))),
\end{equation}%
what implies the relation%
\begin{equation}
\Delta (g)\mathcal{R}_{21}-\mathcal{R}_{21}\Delta _{21}(g)=0\,.  \label{kop}
\end{equation}%
It is well-known (see e.g \cite{maj}) that for any quasitriangular deformed
Poincare algebra the relations (\ref{kop}) follow from the definition of
universal $\mathcal{R}$-matrix%
\begin{equation}
\Delta _{21}(g)=\mathcal{R}\Delta (g)\mathcal{R}^{-1}.
\end{equation}

\section{Covariant braided field commutator and braided field oscillators
algebra in twist deformed QFT}

In twist-deformed quantum field theory the multiplication prescription (\ref%
{ggn}) is determined if \ we know the twist factor $\mathcal{F=F}%
_{(1)}\otimes \mathcal{F}_{(2)}$ and the braid $\Psi _{\mathcal{M},\mathcal{H%
}}$ (see (\ref{fun})). The explicit form of the product of quantum free
fields on noncommutative space-time has therefore a form (we denote $%
\mathcal{F}^{-1}\mathcal{=}\overline{\mathcal{F}}_{(1)}\otimes \overline{%
\mathcal{F}}_{(2)}$) 
\begin{align}
\widehat{\phi }(\widehat{x})\bullet \widehat{\phi }(\widehat{y})& =m_{%
\mathcal{M}\underline{\otimes }\mathcal{H}}[\widehat{\phi }(\widehat{x}%
)\otimes \widehat{\phi }(\widehat{y})]  \label{aaaa} \\
& =\frac{1}{(2\pi )^{8}}\int d^{4}p\int d^{4}q\delta (p^{2}-m^{2})\delta
(p^{2}-m^{2})  \notag \\
& \qquad \qquad \qquad e^{ip\widehat{x}}(\mathcal{R}_{(2)}\triangleright
e^{iq\widehat{y}})\otimes (\mathcal{R}_{(1)}\blacktriangleright A(p))A(q), 
\notag \\
& \overset{W}{\simeq }\frac{1}{(2\pi )^{8}}\int d^{4}p\int d^{4}q\delta
(p^{2}-m^{2})\delta (p^{2}-m^{2})  \notag \\
& \qquad \qquad \qquad (\overline{\mathcal{F}}_{(1)}\triangleright e^{ipx})(%
\overline{\mathcal{F}}_{(2)}\mathcal{R}_{(2)}\triangleright e^{iqy})(%
\mathcal{R}_{(1)}\blacktriangleright A(p))A(q),  \notag
\end{align}%
where by the notation $\overset{W}{\simeq }$ we denote the Weyl homomorphism
in $\widehat{\mathcal{M}}^{(2)}$ with the suitably $\star $-product which
represents the product $e^{ip\widehat{x}}e^{iq\widehat{y}}$ in terms of
classical Fourier exponentials. The actions of free Poincare generators in $%
\overline{\mathcal{F}}_{(1)},\overline{\mathcal{F}}_{(2)}$ on the classical
plane waves are described by the differential realization (\ref{rd1}) and on
the deformed field oscillators the Poincare generators act by the quantum
adjoint action (\ref{adds}).

By using (\ref{dfd}) and (\ref{aaaa}) we shall calculate explicitly the
braided commutator (\ref{cdc1}). We get 
\begin{align}
\lbrack \widehat{\phi }(\widehat{x}),\widehat{\phi }(\widehat{y})]_{\bullet
}^{BR}& =\widehat{\phi }(\widehat{x})\bullet \widehat{\phi }(\widehat{y})-%
\mathcal{R}_{21}\blacktriangleright (\widehat{\phi }(\widehat{y})\bullet 
\widehat{\phi }(\widehat{x}))  \label{51} \\
& \overset{W}{\simeq }\frac{1}{(2\pi )^{8}}\int d^{4}p\int d^{4}q\delta
(p^{2}-m^{2})\delta (q^{2}-m^{2})  \notag \\
& \qquad \lbrack (\overline{\mathcal{F}}_{(1)}\triangleright e^{ipx})(%
\overline{\mathcal{F}}_{(2)}\mathcal{R}_{(2)}\triangleright e^{iqy})(%
\mathcal{R}_{(1)}\blacktriangleright A(p))A(q)  \notag \\
& \qquad -(\overline{\mathcal{F}}_{(1)}\triangleright e^{iqy})(\overline{%
\mathcal{F}}_{(2)}\mathcal{R}_{(2)}\triangleright e^{ipx})(\mathcal{R}_{(1)}%
\mathcal{R}_{(2)}\blacktriangleright A(q))(\mathcal{R}_{(1)}%
\blacktriangleright A(p)).  \notag
\end{align}%
Further in order to obtain explicit formulae we shall consider the canonical
deformation described by twist (\ref{twist0}). As follows from (\ref{rfr})
and (\ref{twist0}) $\mathcal{R}_{21}$ depends only on the fourmomentum
generators $P_{\mu }$ actions given by the formulae%
\begin{equation}
P_{\mu }\rhd e^{ipx}=p_{\mu }e^{ipx}\,,\qquad P_{\mu }\blacktriangleright
A(p)=-p_{\mu }A(p).\,
\end{equation}%
In canonically deformed case we get%
\begin{eqnarray}
\mathcal{R}_{21}\rhd \lbrack e^{ipx}\otimes e^{iqy}] &=&e^{\,i\,\theta ^{\mu
\nu }\,p_{\mu }q_{\nu }}\,\,e^{ipx}\otimes e^{iqy}, \\
\mathcal{R}_{21}\blacktriangleright \lbrack A(p)\otimes A(q)]
&=&e^{\,i\,\theta ^{\mu \nu }\,p_{\mu }q_{\nu }}\,A(p)\otimes A(q)\,,  \notag
\end{eqnarray}%
and the braid $\Psi _{\mathcal{M},\mathcal{H}}$ has the explicit form 
\begin{equation}
\Psi _{\mathcal{M},\mathcal{H}}[A(q)\otimes e^{ipx}]=\mathcal{R}_{(2)}\rhd
e^{ipx}\otimes \mathcal{R}_{(1)}\blacktriangleright A(q)=e^{\,-i\,\theta
^{\mu \nu }\,p_{\mu }q_{\nu }}\,\,e^{ipx}\otimes A(q).\,
\end{equation}

In order to obtain $c$-number braided field commutator one should be able to
factor out in braided field commutator (\ref{cdc1}) the binary relations
satisfied by the field oscillators which describe the field oscillators
algebra. If we use the formula (\ref{fun}) the required factorization in the
braided commutator (\ref{cdc1}) is achieved by the formula\ 
\begin{align}
\lbrack \widehat{\phi }(\widehat{x}),\widehat{\phi }(\widehat{y})]_{\bullet
}^{BR}& \overset{W}{\simeq }\frac{1}{(2\pi )^{8}}\int d^{4}p\int
d^{4}q\delta (p^{2}-m^{2})\delta (q^{2}-m^{2})e^{ipx}e^{iqy}  \label{ccom} \\
& \qquad \qquad \lbrack A(p)\star _{\mathcal{H}}A(q)-\mathcal{R}%
_{21}\blacktriangleright (A(q)\star _{\mathcal{H}}A(p))],  \notag
\end{align}%
where the new multiplication describing the binary oscillator algebra
relation is the following%
\begin{equation}
A(p)\star _{\mathcal{H}}A(q)=m\circ \mathcal{F}\blacktriangleright \lbrack
A(p)\otimes A(q)]=e^{\,\frac{i\,}{2}\theta ^{\mu \nu }\,p_{\mu }q_{\nu
}}A(p)A(q).  \label{nsm}
\end{equation}%
We point out here that the multiplication $\star _{\mathcal{H}}$ is an
inverse of the $\star -$multiplication ($\mathcal{F}^{-1}$ in (\ref{ccdf}))
is replaced in (\ref{nsm}) by $\mathcal{F}$) but it is known from the
literature (e.g. it was used in \cite{wess}) as describing deformed
oscillators algebra.

The following modification of standard free field oscillators algebra (\ref%
{sssa}) describes the binary relation for deformed field oscillators 
\begin{equation}
\delta (p^{2}-m^{2})\delta (q^{2}-m^{2})[A(p)\star _{\mathcal{H}}A(q)-%
\mathcal{R}_{21}\blacktriangleright (A(q)\star _{\mathcal{H}}A(p))]=\epsilon
(p_{0})\delta (p^{2}-m^{2})\delta ^{(4)}(p+q).  \label{ggh}
\end{equation}
The choice (\ref{ggh}) leads to desired properties, namely:

\begin{enumerate}
\item due to the presence of braid factor $\mathcal{R}_{21}$ the deformed
oscillators algebra is covariant under quantum Poincare symmetries

\item it leads to $c$-number value of the braided commutator (\ref{ccom})
\end{enumerate}

If we substitute (\ref{ggh}) into (\ref{ccom}) we obtain the final formula%
\begin{equation}
\lbrack \widehat{\phi }(\widehat{x}),\widehat{\phi }(\widehat{y})]_{\bullet
}^{BR}\overset{W}{\simeq }\Delta (x-y;m^{2}),  \label{loc}
\end{equation}%
with the braided commutator for canonically deformed free quantum fields
given by known standard Pauli-Jordan function (see (\ref{spjs})) .

It should be noted that the choices of $\star -$multiplication (see (\ref%
{ccdf})) and of the covariant braid $\Psi _{\mathcal{M},\mathcal{H}}$ (see (%
\ref{fun}) are necessary for getting the twist-covariant algebra of deformed
field operators. The braid factor $\mathcal{R}_{21}$ which is an intertwiner
in quantum quasitriangular Poincare-Hopf algebra appears in our framework on
three levels:

\begin{enumerate}
\item in the Weyl realization of the algebra $\widehat{\mathcal{M}}$ as
expressing the "braided commutativity" of the algebra of classical fields on
Minkowski space-time with the $\star -$multiplication (see e.g. \cite{alv}).

\item in the algebra $\widehat{\mathcal{H}}$ (see (\ref{ggh})) of quantized
field oscillators which is covariant under the action $\blacktriangleright $
of the Poincare symmetry generators $g$.

\item in the algebra $\Phi (\widehat{\phi }(\widehat{x}),\bullet )$ of
deformed NC free quantum field $\widehat{\phi }(\widehat{x})$ (see (\ref{fie}%
)) firstly in the definition of multiplication $m_{\mathcal{M}\otimes 
\mathcal{H}}$ (see (\ref{ggn})), and further in the definition of braided
field commutator (\ref{cdc1}).
\end{enumerate}

We point out that the nonstandard multiplication of the deformed oscillators
given by (\ref{nsm}) and the relation (\ref{ggh}) are selected by the
requirement of braided $\bullet -$locality of NC quantum free fields or
equivalently by the $c-$number value of braided field commutator (see (\ref%
{loc})). In NC QFT covariant under quantum Poincare symmetries the standard
locality condition is therefore modifed not only by the introduction of $%
\star -$multiplication of fields on Minkowski space but also by the
deformation of field commutator into the braided one. Let us observe that in
the example of canonical twist (\ref{twist0}) the rhs of (\ref{51}) and (\ref%
{ccom}) describing the braided commutator of fields $\widehat{\phi }(x),%
\widehat{\phi }(y)$ after the Weyl map vanishes if the points $x,y$ are
separated in space-like way $((x-y)^{2}<0)$. We see that the modification of
locality in traditional approach (see (\ref{fd})) is replaced by the
following braided $\bullet -$locality relation which should be understood
after performing the Weyl map which introduces classical space-time points $%
x,y$\footnote{%
If we introduce the extension of the $\star $-product on $\mathcal{M}$ into
the symbol $\circledast $ (see footnote g), the braided $\star $-locality
can be rewritten as $[\widehat{\phi }(\widehat{x}),\widehat{\phi }(\widehat{y%
})]_{\circledast }=0$ for $(x-y)^{2}<0$ where $[A,B]_{\circledast
}=A\circledast B-B\circledast A$.} 
\begin{equation}
\lbrack \widehat{\varphi }(x),\widehat{\varphi }(y)]=0\overset{\text{quantum}%
}{\underset{\text{deformation}}{\Longrightarrow }}[\widehat{\phi }(\widehat{x%
}),\widehat{\phi }(\widehat{y})]_{\bullet }^{BR}\overset{W}{\simeq }0.
\label{ree}
\end{equation}

The formula (\ref{loc}) provides an explicit example of NC quantum field
satisfying the braided $\bullet -$locality condition (\ref{ree}).

\section{Conclusions}

In this paper we present the quantum-covariant braided formulation of the
theory of noncommutative quantum free fields. We restricted our
considerations to binary products of such fields. For twist-deformed
noncommutative fields the extension of our formalism to n-ary associative
products is rather straightforward, with the associativity of braided field
products following from the hexagon relation satisfied by braid $\mathcal{R}%
_{21}$ (see e.g. \cite{maj}).

From our quantum covariance requirements we obtained the braided form (\ref%
{dfd}) of \ deformed field commutator, the braided form (\ref{bra}) of
deformed oscillator algebra and braided $\bullet -$multiplication (\ref{ggn}%
) in the algebra of deformed quantum free fields. We add that the elements
of our construction were present in previous papers \cite{wess},\cite{zahn},%
\cite{fi},\cite{fi2}. The closest to our consideration is the approach of
Fiore \cite{fi},\cite{fi2}\ where the braided commutator is however defined
with braid factor $\mathcal{R}_{21}$ acting on fields $\widehat{\phi }(%
\widehat{x})$ by the action $\trianglerighteq $, described by (\ref{nn1})
(this was already implicit in the second option of \cite{wess}, see formula
(46) there). In such a case the braid factor becomes trivial, because due to
the relations $\varepsilon (\mathcal{R}_{(1)})\otimes \mathcal{R}_{(2)}=%
\mathcal{R}_{(1)}\otimes \varepsilon (\mathcal{R}_{(2)})=\mathbf{1}\otimes 
\mathbf{1}$ ($\varepsilon $ is the counit) and (\ref{nn2}) we get $\mathcal{R%
}_{(1)}\trianglerighteq \widehat{{\phi }}(\widehat{x})\otimes \mathcal{R}%
_{(2)}=\widehat{{\phi }}(\widehat{x})\otimes \mathbf{1}$, \ $\mathcal{R}%
_{(2)}\trianglerighteq \widehat{{\phi }}(\widehat{x})\otimes \mathcal{R}%
_{(1)}=\widehat{{\phi }}(\widehat{x})\otimes \mathbf{1},$ whence 
\begin{equation}
g\trianglerighteq (\widehat{{\phi }}(\widehat{y})\widehat{{\phi }}(\widehat{x%
}))=\widehat{{\phi }}(\widehat{y})\widehat{{\phi }}(\widehat{x}),
\end{equation}%
and the braided field commutator becomes a standard one. The latter also
gives the $c$-number function at the rhs of (\ref{loc}), which fulfills (\ref%
{ree}) at spacelike distances.

Our explicit calculations have been given for the simplest case of canonical
twist deformation. If however the twist factor depends as well on the
Lorentz generators $M_{\mu \nu }$ (see e.g. \cite{lw}), the formulae
describing the algebra of deformed quantum fields are more complicated. In
such a case after the Weyl map the bidifferential operator describing $\star
-$product in algebra $\widehat{\mathcal{M}}$ depends also on the space-time
coordinate $x_{\mu }$ and explicit calculations are much more complicated.
However, in principle the presented here braided fields approach can be
applied to general quasitriangular quantum deformation of free quantum
fields.

\bigskip

\textbf{Acknowledgements: }

\bigskip We would like to thank Gaetano Fiore for reading the paper and
providing important comments. JL is thankfull to Goran Djokovic and Jelena
Stankovic for the invitation to participate in Memorial Julius Wess 2011
Workshop. The paper was supported by Polish NCN grant 2011/01/B/ST2/03354.

\end{document}